\documentclass[a4paper,fleqn,usenatbib]{mnras}

\usepackage[normalem]{ulem}

\usepackage{newtxtext,newtxmath}
\usepackage{siunitx}
\usepackage[T1]{fontenc}
\usepackage{ae,aecompl}

\usepackage{graphicx}	
\usepackage{amsmath}	
\usepackage{pdflscape}
\usepackage{hyperref}

\defcitealias{2000PASP..112.1360A}{AK00}

\title[M92 with NIRCam@\textit{JWST}]{Photometry and astrometry with \textit{JWST} -- I. NIRCam
   Point Spread Functions and the first \textit{JWST} colour-magnitude
  diagrams of a globular cluster}

\author[D.\ Nardiello]{D.\ Nardiello$^{1,2}$\thanks{E-mail: domenico.nardiello@inaf.it}, 
L.\ R.\ Bedin$^{1}$, 
 A.\ Burgasser$^{3}$,
M.\ Salaris$^{4,5}$,
S.\ Cassisi$^{5,6}$,
M.\ Griggio$^{7,1}$,
M.\ Scalco$^{7,1}$ \\
$^{1}$Istituto Nazionale di Astrofisica - Osservatorio Astronomico di Padova, Vicolo dell'Osservatorio 5, IT-35122, Padova, Italy \\
$^{2}$Aix Marseille Univ, CNRS, CNES, LAM, Marseille, France \\
$^{3}$Center for Astrophysics and Space Sciences (CASS), University of California, San Diego, La Jolla, CA, 92093, USA\\
$^{4}$Astrophysics Research Institute, Liverpool John Moores University, 146 Brownlow Hill, Liverpool L3 5RF, UK \\
$^{5}$Istituto Nazionale di Astrofisica - Osservatorio Astronomico di Abruzzo, Via M. Maggini, I-64100, Teramo, Italy \\
$^{6}$INFN - Sezione di Pisa, Largo Pontecorvo 3, I-56127 Pisa, Italy  \\
$^{7}$Dipartimento di Fisica, Universit\`a di Ferrara, Via Giuseppe Saragat 1, I-44122, Ferrara, Italy\\
}

\date{Accepted 2022 September 13. Received 2022 September 13; in original form 2022 August 16}

\pubyear{2022}
\begin{document}
\label{firstpage}
\pagerange{\pageref{firstpage}--\pageref{lastpage}}
\maketitle

\begin{abstract}
As the \textit{James Webb Space Telescope (\textit{JWST})} has become
fully operational, early-release data are now available to begin
building the tools and calibrations for precision point-source
photometry and astrometry in crowded cluster environments.
Here, we present our independent reduction of NIRCam imaging of the
metal-poor globular cluster M~92, which were collected under
Director’s Discretionary Early Release Science programme ERS-1334.
We derived empirical models of the Point Spread Function (PSF) for
filters F090W, F150W, F277W, and F444W, and find that these PSFs: (i)
are generally under-sampled (FWHM$\sim 2$~pixel) in F150W and F444W and
severely under-sampled (FWHM$\sim 1$ pixel) in F090W and F277W; (ii)
have significant variation across the field of view, up to $\sim
15$--20~\%; and (iii) have temporal variations of $\sim$ 3--4~\%
across multi-epoch exposures.
We deployed our PSFs to determine the photometric precision of NIRCam
for stars in the crowded, central regions of M~92, measured to be at
the $\sim$0.01\,mag level.
We use these data to construct the first \textit{JWST}
colour-magnitude diagrams of a globular cluster. Employing existing
stellar models, we find that the data reach almost the bottom of the
M~92 main sequence ($\sim$0.1~$M_{\odot}$), and reveal 24 white dwarf
candidate members of M~92 in the brightest portion of the white dwarf
cooling sequence. The latter are confirmed through a cross-match with
archival {\it HST} UV and optical data.
We also detect the presence of multiple stellar populations along the
low-mass main sequence of M~92.
%

\end{abstract}




\section{Introduction}
In the past 30 years, space telescopes have revolutionised the world
of astronomy, with new and, sometimes, unexpected discoveries.
Notable examples include \textit{Hipparcos}
(\citealt{2000A&A...355L..27H}) and \textit{Gaia}
(\citealt{2016A&A...595A...1G}), which significantly advanced the
field of astrometry and contributed to our understanding of stellar
astrophysics and Milky Way structure; \textit{Kepler}
(\citealt{2010Sci...327..977B}) and \textit{TESS}
(\citealt{2015JATIS...1a4003R}), whose precision light curves greatly
expanded the fields of exoplanets and variable stars; \textit{Spitzer}
(\citealt{2004ApJS..154....1W}), \textit{Herschel}
(\citealt{2010A&A...518L...1P}), and \textit{WISE}
(\citealt{2010AJ....140.1868W}), whose sensitivity at infrared (IR)
wavelengths advanced studies of star forming regions, distant
galaxies, and cosmology; and \textit{Chandra}
(\citealt{2002PASP..114....1W}) and \textit{XMM-Newton}
(\citealt{2001A&A...365L...1J}), which enabled X-ray observations of
the most energetic phenomena in the near and distant Universe.
Arguably the most successful and longest-lived of the space
observatories has been the \textit{Hubble Space Telescope}
(\textit{HST}), whose instruments enable UV, optical, and infrared
imaging and spectroscopy at high spatial resolution and sensitivity,
and continues to provide revolutionary results in all areas of
astrophysics.
\begin{figure*}
\includegraphics[width=0.95\textwidth]{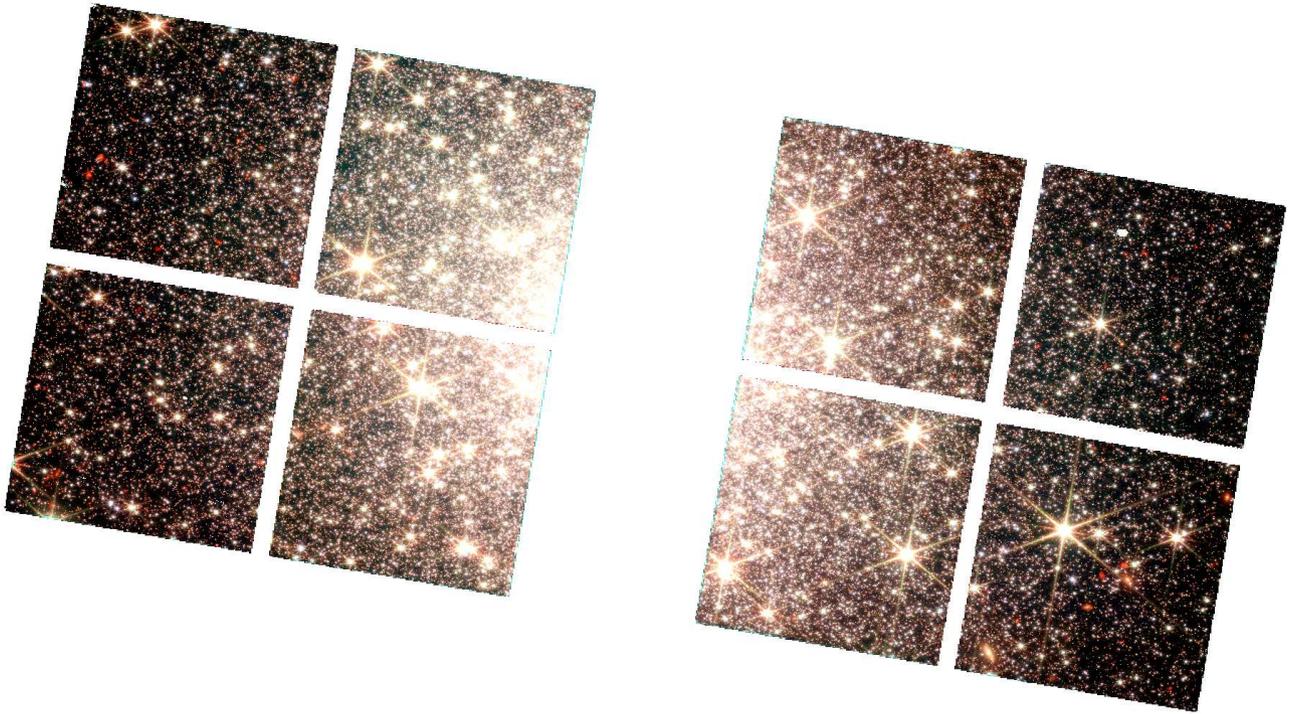}
\caption{Three-colour stacked image of the NIRCam field of view
  centred on M~92. The filters F444W, F150W, and F090W have been used
  for the red, green, and blue channels, respectively. The field is
  oriented with North up and East left. In the F444W stacked image, we
  masked pixels corresponding to the gaps of the SW
  channel. \label{fig:0}}
\end{figure*}

On 25 December 2021, the NASA/ESA/CSA \textit{James Webb Space
  Telescope} (\textit{JWST}, \citealt{2006SSRv..123..485G}) was
successfully launched on an Ariane 5 rocket from the Guiana Space
Centre in Kourou, French Guiana.  Over the subsequent six months after
arrival, \textit{JWST} successfully passed all phases of commissioning
and testing, and on 11 July 2022 the first science images from the
facility were made public.  \textit{JWST}'s 6.5~m primary mirror is
formed by 18 hexagonal mirror segments made of gold-plated beryllium,
which allow the telescope to observe at both near-IR and mid-IR
wavelengths (0.6-28.3 $\mu$m). Four science instruments (NIRCam,
NIRSpec, MIRI, NIRISS) and a guide camera (FGS) are mounted in the
Integrated Science Instrument Module. Their varied capabilities are
enabling analysis of the light of the first galaxies, the study of
exoplanet atmospheres, and improved understanding of the origins and
evolution of resolved stellar populations in clusters and nearby
galaxies (\citealt{2019arXiv190308670G, 2019BAAS...51c.554M,
  2019BAAS...51c..45R,2019BAAS...51c.449W}).

Before the launch of \textit{JWST}, 13 Director's Discretionary Early
Release Science (DD-ERS)
programmes\footnote{https://www.stsci.edu/jwst/science-execution/approved-ers-programs}
were selected to conduct the first scientific observations with the
facility after the completion of commissioning.  These data are
immediately released to the public.

This is the first in a series of papers aimed at exploring the
astrometric and photometric capabilities of \textit{JWST}'s
instruments based on data collected during the DD-ERS programmes, with
the goal of developing the tools needed to obtain high-precision
photometry and astrometry of stars in crowded environments.  In this
work, we extract effective Point Spread Functions (PSFs) and
photometry for stars in the very metal-poor globular cluster NGC~6341,
based on images collected with the Near Infrared Camera (NIRCam) of
\textit{JWST} as part of programme ERS-1334 (PI: Weisz).  Subsequent
works will focus on correcting geometric distortion and obtaining
precision astrometry of stars with NIRCam (Paper II, Griggio et al. in
preparation), analysis of resolved galaxies (Paper III, Bedin et
al. in preparation), and on the use of the NIRISS and FGS images for
crowded field studies (Paper IV, Nardiello et al. in preparation).
This article is structured as follows: Section~\ref{sec:obs} describes
the NIRCam observations presented in this work.  Section~\ref{sec:psf}
reviews the procedures used to derive effective PSF models for each of
the four NIRCam filters used, and to extract the first \textit{JWST}
photometric catalogues of stars in M~92.  Section~\ref{sec:cal} reports
the procedures used to transform instrumental magnitudes into a
photometric system defined by theoretical models.
Section~\ref{sec:cmd} describes an analysis of the M~92 colour-magnitude diagrams (CMDs) obtained
by combining \textit{JWST} and {\it HST} imaging data.
Section~\ref{sec:sum} summarises our conclusions.

\begin{figure*}
\includegraphics[width=0.95\textwidth]{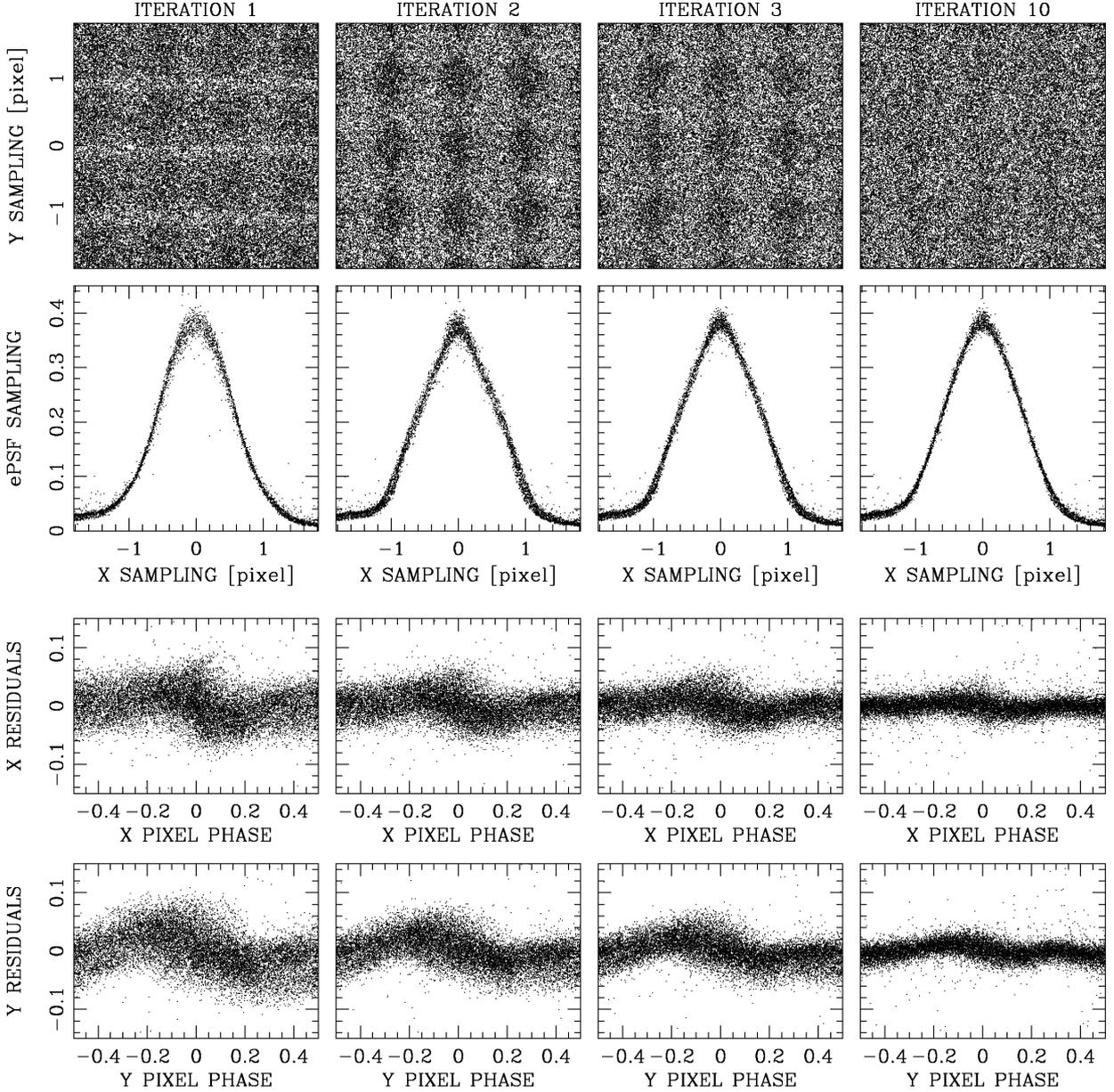}
\caption{ Overview of the iterative process of determine the ePSF
  model for filter F090W for detector 4 of module A, in the case of no
  spatial variation of the PSF across the field of view.  The top row
  shows the location of the estimated value of the ePSF with respect
  to the centre of the stars (0,0), i.e. the ePSF $x$/$y$ sampling;
  only 2.5\% of the points are plotted.  The second row displays
  the ePSF $x$ sampling for a slice across the centre of the ePSF with
  $|\delta y|<0.05$.  The bottom two rows illustrate the pixel-phase
  errors along the $x$ and $y$ axes, respectively.  From left to
  right, the panels correspond to iterations 1, 2, 3, and 10.
  \label{fig:1}}
\end{figure*}

\section{Observations}
\label{sec:obs}
The NIRCam imaging camera covers the red optical and near-IR
wavelength range 0.6-5 $\mu$m in two simultaneous channels: the {short
  wavelength} (SW) channel (0.6-2.3 $\mu$m) and the {long wavelength}
(LW) channel (2.4-5.0 $\mu$m).  Each channel consists of two modules
(A and B) that operate in parallel.  The field of view of each module
is 2$\farcm$2 $\times$ 2$\farcm$2 (4.84~arcmin$^2$), and these are
separated by a gap of $\sim 44$\,arcsec. The total field of view for
each channel is about 9.7~arcmin$^2$.  In the SW channel, gaps of
$\sim 5$\,arcsec separate the four detectors that constitute each
module.  Both modules in each channel operate with the same set of
narrow-, medium-, and wide-band filters.  The imaging resolution for
the SW channel is $\sim 31$ mas/pixel, while for the LW channel it is
$\sim 63$~mas/pixel.

DD-ERS programme ERS-1334 used NIRCam to image NGC~6341 (M~92), a very
metal-poor globular cluster ([Fe/H]=$-$2.3,
\citealt{1996AJ....112.1487H}) located at a distance of $\sim 8.5$~kpc
from the Sun.  Observations were obtained over 20-21 June 2022 for a
total 1.95 hours, using filters F090W ($\lambda \sim 0.8$--1.0$\mu$m)
and F150W ($\lambda \sim 1.3$--1.7$\mu$m) in the SW channel, and F277W
($\lambda \sim 2.4$--3.1$\mu$m) and F444W ($\lambda \sim
3.9$--5.0$\mu$m) in the LW channel.  For each filter, four exposures
of 311.37~s each were obtained using the , \texttt{SHALLOW4} readout
mode (4 frames averaged and 1 frame skipped), for a total integration
time of 1245.48~s per filter. These images were obtained in a 4-point
sub-pixel dither pattern to mitigate cosmic rays, bad pixels, and
improve the sampling of the PSF.  Both NIRCam channels were centred
such that the the 44~arcsec gap between the modules covered the centre
of the cluster, as illustrated in the stacked, three-colour (F090W,
F150W, F444W) image shown in Figure~\ref{fig:0}.\\

For our analysis, we used the single exposure calibrated images
(\texttt{\_cal}) that are created by the Stage 2 pipeline
\texttt{calwebb\_image2}\footnote{\url{https://jwst-pipeline.readthedocs.io/}}.
Each pixel is calibrated by the pipeline to units of MJy/steradian. We
converted these values into counts by using the header keywords
\texttt{PHOTMJSR} (flux density in MJy/steradian corresponding to 1
count/second) and \texttt{XPOSURE} (effective exposure time).  We also
used the Data Quality (DQ) image included in the \texttt{\_cal} fits
data cube to flag bad and saturated pixels on the science image.

A total of $8 \times 2048\,{\rm pixel} \times 2048\,{\rm pixel}$
images are associated to each exposure obtained with the SW channel
(four images for each module), while $2 \times 2048\,{\rm pixel}
\times 2048\,{\rm pixel}$ images are associated to each exposure
obtained with the LW channel (one for each module).
We excluded from our reduction and analysis the \texttt{\_cal}
exposures with root-name \texttt{jw01334001001\_02101\_00003}
(involving filters F090W and F277W, for a total of 10 images) as they
were found to be unusable.
We therefore reduced 24, 32, 6, and 8 images in F090W, F150W, F277W,
and F444W filters, respectively.

\section{PSF modelling and data reduction}
\label{sec:psf}
As reported in the \textit{JWST}
documentation\footnote{\url{https://jwst-docs.stsci.edu/}}, the NIRCam
PSFs (in imaging mode) are Nyquist sampled for wavelengths longer than
2.0\,$\mu$m in the SW channel and 4.0\,$\mu$m in the LW channel,
corresponding to full-width at half-maximum FWHM $\sim 2$--3~pixels.
At shorter wavelengths, the PSFs are under-sampled.
Deriving a correct model for an under-sampled PSF is a challenging but
necessary task, as incorrect PSFs can introduce systematic errors in
the extraction of positions and fluxes of stars
(\citealt{2000PASP..112.1360A}).

We followed the empirical approach developed by \citet[hereafter
  AK00]{2000PASP..112.1360A} for the {\it HST} Wide-Field Planetary
Camera 2 (WFPC2) to obtain a model of the \textit{effective}
pixel-convolved PSF (ePSF). This approach has been used to derive
``library'' (reference) ePSFs for the {\it HST} Advanced Camera for
Survey (ACS, \citealt{2004acs..rept...15A,2006acs..rept....1A}) and
the Wide Field Camera 3
(WFC3,\citealt{2015wfc..rept....8A,2016wfc..rept...12A}), which to
this day represent {the state of the art} for point-source photometry
and astrometry. The ePSF approach has also been applied to the
modelling of under-sampled PSFs in Kepler/K2 images
(\citealt{2016MNRAS.456.1137L,2016MNRAS.463.1780L,2016MNRAS.463.1831N}).
The procedure outlined below was applied independently for each set of
images in a given detector/filter combination.

To break the degeneracy between position and flux of a source (see
Fig.~1 of \citetalias{2000PASP..112.1360A}) to derive a well-sampled
PSF model, it is necessary to precisely constrain the positions and
fluxes for a set of stars.
However, this information can not be obtained accurately if the PSF is
not appropriately modelled. As such, an iterative approach is
required: starting from a first-guess catalogue of positions and
fluxes of isolated, bright, unsaturated point sources, we alternately
derive the ePSF model and the stars' positions and fluxes, improving
at each iteration both sets of quantities until reasonable convergence
is achieved.

\subsection{A first-guess catalogue of stars}
\label{sect:mastercat}

For each image, we first extracted a initial estimate of the empirical
PSF following the approach by \citet{2006A&A...454.1029A}.  PSFs
models are obtained for each image in a total empirical way, by
normalising a sample of isolated, bright stars for their total flux,
and averaging them on a grid of $201\times201$ points defined in a
space that super-samples the PSF pixel by a factor 4.  These empirical
PSFs, even if already good enough for measuring with a first
approximation the positions and the fluxes of the stars, are not
sufficient for high-precision photometric and astrometric measurements
of the stars, particularly in crowded environments.
We then used these initial PSFs to measure the positions and fluxes of
isolated (the peak must be at least 10 pixels from the nearest peak),
bright (the minimum flux above the local sky must be at least 3000
counts), unsaturated stars in each image.
In this step we used the software \texttt{img2xym}, developed by
\citet{2006A&A...454.1029A} for the Wide Field Imager (WFI) @ ESO/MPG
2.2m telescope, and adapted to NIRCam data. This algorithm finds,
models, measures, and subtracts stars that are progressively fainter
through an iterative process, simultaneously fitting the positions and
fluxes of each star and its (potentially overlapping) neighbours.
We cross-identified the stars in each image in a given detector/filter
dither set, then transformed the positions and fluxes of the stars
into a single catalogue anchored to a common reference system
associated to the first image in each set, using six-parameter linear
transformations and photometric zero-point shifts. The transformed
positions and fluxes of each star detected in at least three images
are then used to define a master catalogue that serves as the
reference for modelling the ePSF model for a given detector and filter.
Each catalogue contains between 1500 and 6000 stars depending on the
channel, module, detector, and filter it represents.

\begin{figure*}
\includegraphics[bb=80 280 486 676, width=0.45\textwidth]{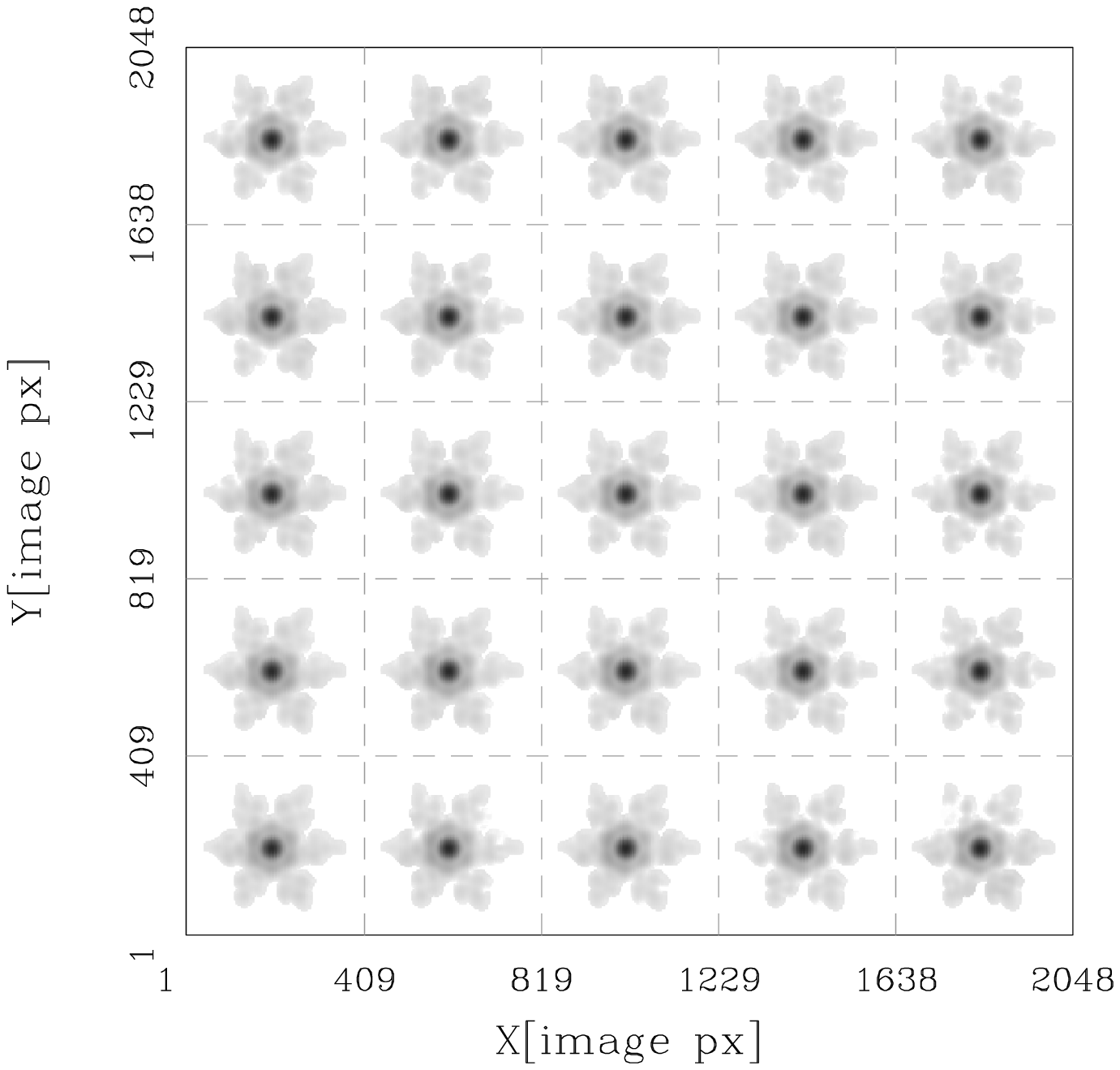}
\includegraphics[width=0.45\textwidth]{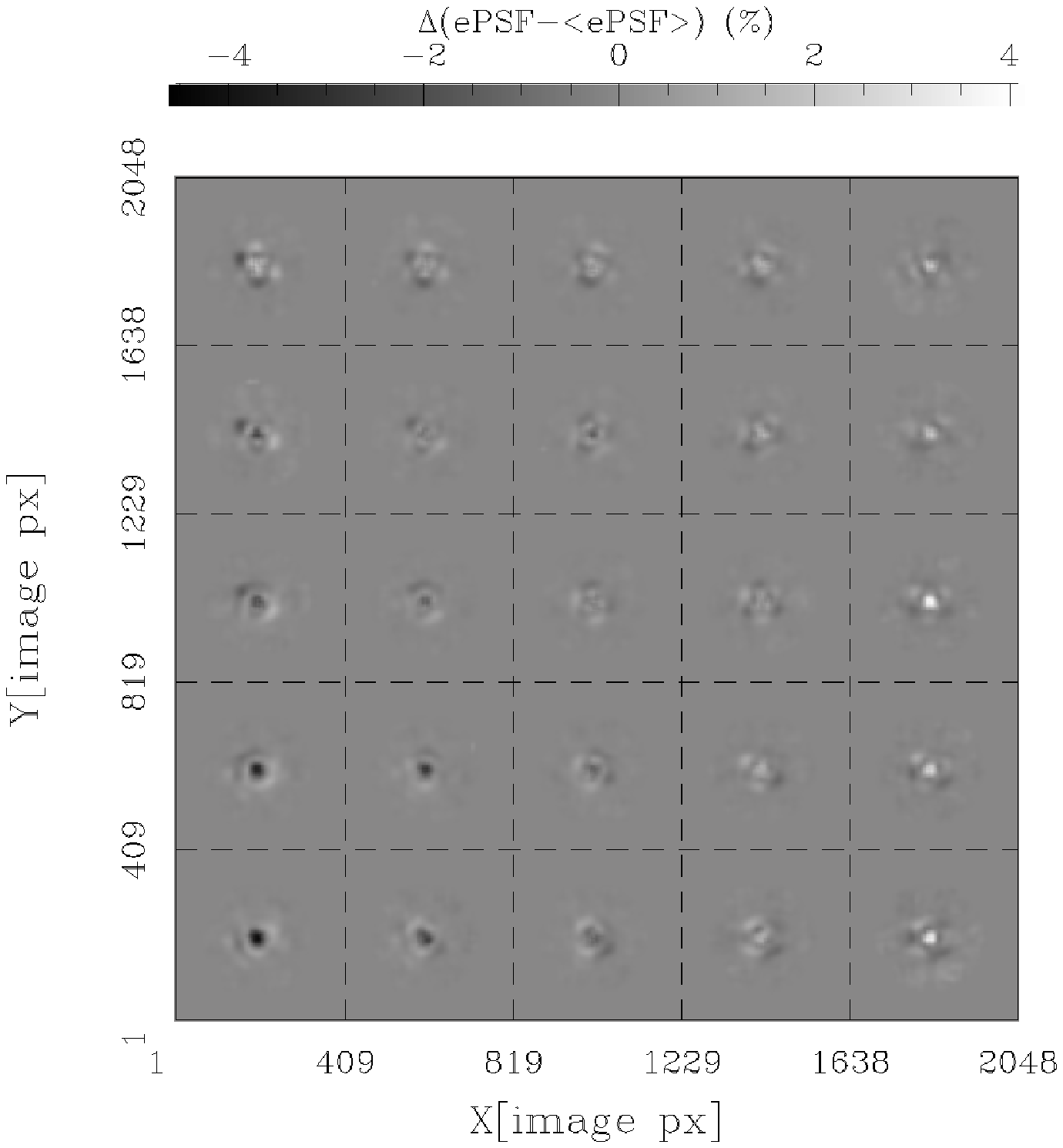}
\caption{ Overview of the spatial variation of the ePSF model for
  filter F150W and detector 1 of module A in the SW channel. The left
  panel shows the final $5 \times 5$ array of ePSFs calculated in
  image sub-regions of size 409~pixel $\times$ 409~pixel.  The right
  panel shows the difference between each ePSF and an average of all
  25 ePSFs across the detector. The scale of the variation compared to
  the normalised ePSF is indicated by the colour bar above the panel.
  \label{fig:2}}
\end{figure*}

\subsection{Obtaining the Effective Point Spread Function}

Given an individual star centred at ($x_{\star}, y_{\star}$) and having flux
$f_{\star}$, every pixel ($i,j$) of an image close to the star samples the
normalised ePSF $\psi$ as:
\begin{equation}
  \psi (\Delta x, \Delta y) = \frac{P_{i,j}-s_{\star}}{f_{\star}}
\end{equation}
where $P_{i,j}$ is the count value of the image at pixel ($i$,$j$),
$(\Delta x, \Delta y) = (i- x_{\star}, j- y_{\star})$ is the pixel
offset from the star's centre, and $s_{\star}$ is the local sky
background, measured as the mode of the pixel value distribution
computed in an annulus of radii $r_{\rm in}= 7$~pixel and $r_{\rm
  out}= 13$~pixel and centred on the star.  Knowing the flux and
position of each star in each exposure, each pixel ($i,j$) close to
the star represents a sampling of the PSF at one point in the
two-dimensional array that defines the ePSF.

Our procedure to determine an overall ePSF model followed these steps
(we refer the reader to
  \citetalias{2000PASP..112.1360A} for a more detailed description of
  these steps):
\begin{enumerate}
\item We first transformed the positions of the stars in the master
  catalogue onto the reference system of each individual image using
  the transformations described above.
\item We then converted each pixel value within 10 pixels of the star's centre in a
  given image into an estimate of the corresponding point in the ePSF model.
  Given the large number of stars in each master catalogue,
  there are of order millions of point sampling used to determine the ePSF model.
\item The ePSF model was constructed by projecting these individual
  point samplings from the 10~pixel $\times$ 10~pixel original image
  scale to a finer ePSF grid super-sampled by a factor of 4 (41~points
  $\times$ 41~points).  In the 1st iteration, each ePSF grid-point was
  calculated as the 2.5$\sigma$-clipped average of point sampling
  within a square of 0.25 pixels in $\Delta x$ and $\Delta y$
  coordinates.  In subsequent iterations, we first subtracted from
  each sampling the corresponding value in the current ePSF model,
  then calculated the 2.5$\sigma$-clipped average of the residuals in
  each 0.25~pixel$\times$0.25~pixel grid point. This residual grid was
  then added to the last available ePSF model, and the result was
  smoothed with a combination of quadratic and quartic kernels,
  re-centred, and re-normalised ;
\item Using the updated ePSF model, we remeasured the positions and
  fluxes of stars in the master catalogue in each image; then
  transformed the positions back to the original reference frame and
  computed average fluxes to update the catalogue.
\end{enumerate}

We performed this iterative process ten times, assuming a single ePSF
model (no spatial variation). Figure~\ref{fig:1} shows the improvement
of the ePSF model from the first iteration to the tenth iteration. In
the first iteration, the point samplings are not homogeneously
distributed, and the ePSF shape has significant scatter among the
samples.  As the ePSF model improves, both the point samplings and
scatter in the ePSF profile improve.  The improvement of the ePSF
model can be particularly discerned in the distribution of pixel phase
errors, i.e. the difference between the expected positions
(transformed from the master catalogue to each individual image) and
the measured positions of the stars (obtained using the ePSF model at
each iteration), as a function of the pixel
phase \footnote{Specifically, the pixel phase of a star, defined as
  $\phi_x = x-{\rm int}(x+0.5)$ and $\phi_y = y-{\rm int}(y+0.5)$, is
  the location of the star with respect to the pixel
  boundaries.}. This distribution shows a sinusoidal pattern in the
first iteration, expect for an offset ePSF model; this pattern
flattens as the ePSF model improves.

After the tenth iteration, we saw no improvement in the pixel phase
error distribution or sample scatter in ePSF model.  However, PSFs
vary with location across the detector due to projection effects, so
subsequent iterations of this process introduced a spatial variation
to the ePSF model. This variation was done by dividing the image into
sub-regions, then calculating the ePSF model in each of these regions
following the steps above, starting with the ``global'' ePSF model
calculated at the end of the tenth iteration.  We increased the number
of sub-regions in stages, starting with 2$\times$2 sub-regions of
dimension 1024~pixel $\times$ 1024~pixel (iterations 11-13), then
3$\times$3 sub-regions of dimension 682~pixel $\times$ 682~pixel
(iterations 14-16), and finally 5$\times$5 sub-regions of dimension
409~pixel $\times$ 409~pixel (iterations 17-21).  Figure~\ref{fig:2}
shows an example of the final ePSF array for filter F150W and detector
1 of module A in the SW channel.  The peak-to-peak variation is about
9~\% in this case, while maximum variations of up to 15--20~\% were
found for the ePSF arrays in F090W images.

\begin{figure*}
\includegraphics[width=0.95\textwidth]{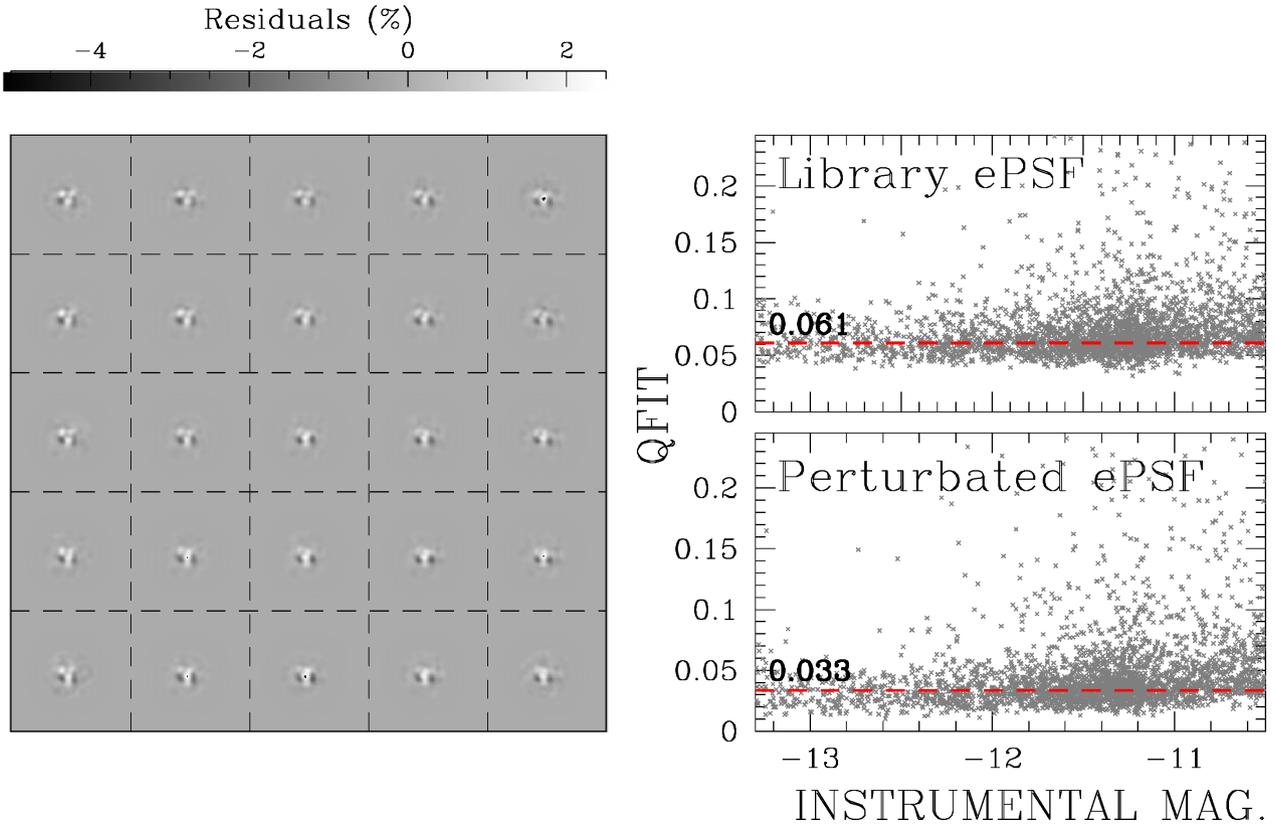}
\caption{Overview on the results obtained from library ePSF
  perturbation from our M~92 ePSF grid to observations of the WLM
  galaxy, also obtained with NIRCam. The left panel shows the
  residuals added to the library ePSFs to obtain the perturbed
  ePSFs. The total variation across the grid goes from $\sim -4\%$ to
  $\sim 2\%$.  The right panel shows the quality-of-fit (qfit) as a
  function of instrumental magnitude before (top panel) and after
  (bottom panel) perturbation of the library ePSF.  The median qfit
  for bright stars (instrumental magnitude between $-13.0$ and
  $-10.5$) declines from $\sim 0.06$ to $\sim 0.03$.
  \label{fig:3}}
\end{figure*}

\begin{figure*}
\includegraphics[bb=26 402 577 698, width=0.95\textwidth]{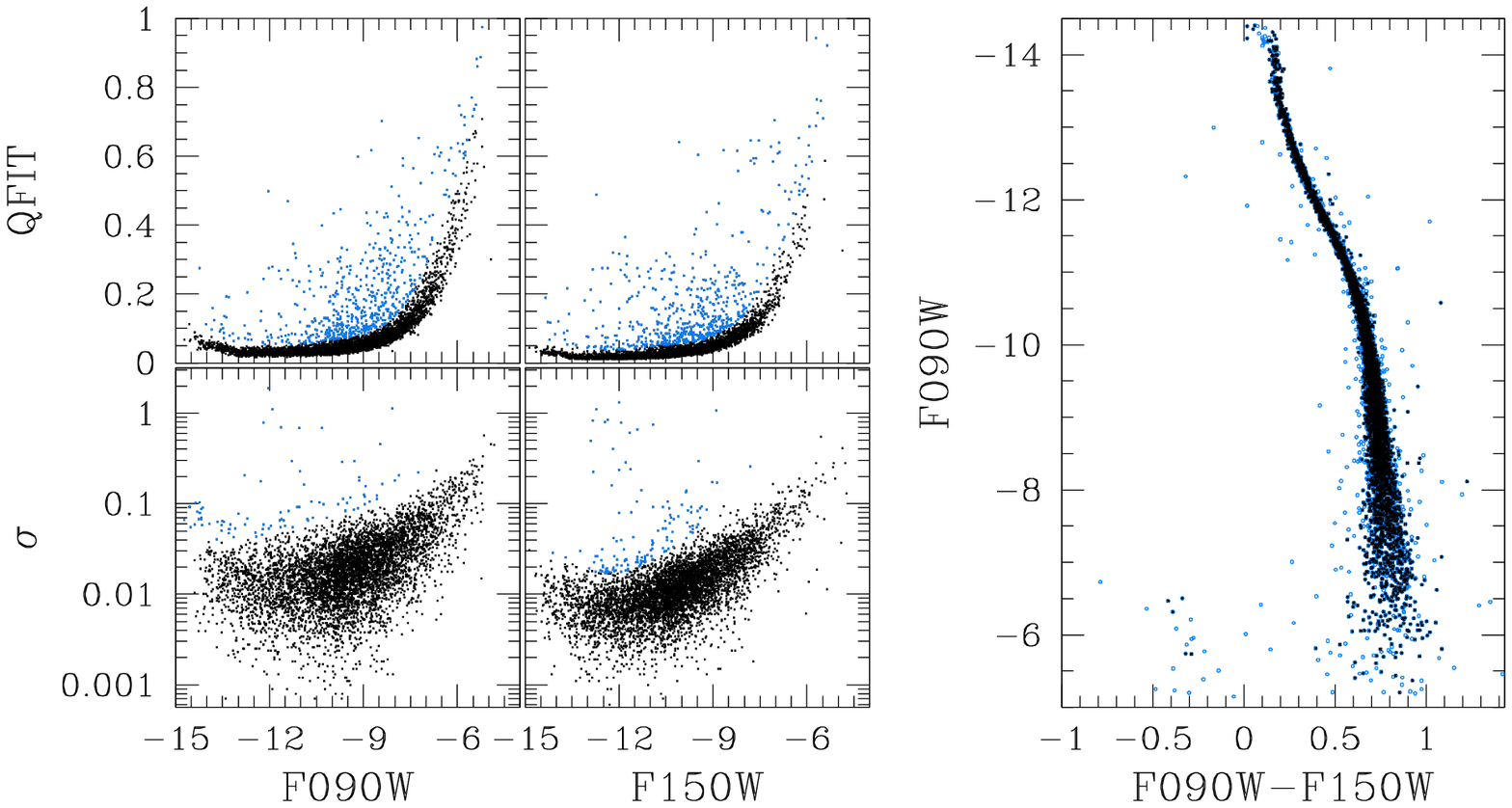}
\caption{Procedure adopted for rejecting bad detection and
  measurements in point-source photometry.  The left panels show the
  selections performed in the photometric error (bottom panels) and
  qfit distributions (top panels) for the F090W (left) and F150W
  (right) filters; azure points indicate rejected sources. The right
  panel shows the instrumental F090W versus F090W-F150W CMD for
  well-measured (black) and rejected sources (azure). This CMD is
  based on the final source catalogues for images obtained with
  detector 1 of module A in the SW channel. \label{fig:4}}
\end{figure*}

\subsection{Time variability of the ePSF}
If the library ePSF model determined above is constant in time, it
could in principle be used to derive the positions and fluxes of stars
in all images obtained with NIRCam. However, past experience with {\it
  HST} shows that, even in ideal and stable conditions, PSFs are known
to vary with time.

In order to measure and minimise the effects of time variation of the
ePSF model, we performed a perturbation analysis originally introduced
in \citet{2006acs..rept....1A} (see also
\citealt{2008AJ....135.2114A}) and deployed in several subsequent {\it
  HST} works (e.g., \citealt{2018MNRAS.481.3382N}).  Briefly, for a
given detector/filter set, we start with the library ePSF measured
from an initial data-set (in this case, the M~92 observations), and
perturb it to account for ePSF variations that emerge in a second
data-set.  For the latter, we used NIRCam observations of the
Wolf-Lundmark-Melotte (WLM) irregular galaxy
(\citealt{1909AN....183..187W, 1926MNRAS..86..636M,
  1950MeLuS.128....5H}), also obtained in programme ERS~1334 about one
month after the M~92 observations.\footnote{We specifically evaluated
  the image \texttt{jw01334005001\_02101\_00001\_nrcb3\_cal},
  corresponding to channel SW, module B, detector 3, and filter
  F090W.}  We follow a similar iterative procedure as outlined above.
We started by using the initial library ePSF model to measure the
positions and fluxes of a set of bright, unsaturated, isolated stars
in the second data-set. We then subtracted the ePSF models from these
stars, and re-sampled the residuals onto a grid of residuals, that will
be used to perturb the original library ePSFs.  The spatial grid of
mean residual can vary from $1 \times 1$ to $5 \times 5$ (same grid as
the ePSFs) based on the number of stars available in the field of view
of the image (minimum 15 stars in a sub-region).  Each point of the
grid corresponds to a point of the image, and, when the used grid is
smaller than $5 \times 5$, we used bi-linear interpolations to calculate
the perturbation to add to each of the 25 library ePSFs.  We repeated
this procedure five times to converge to an optimally perturbed ePSF
model grid.

Figure~\ref{fig:3}, demonstrates the results of the library ePSF
perturbation. The 5$\times$5 residual array between the M~92 ePSF
model and final WLM ePSF model shows variations of order 3--4~\%,
indicating that temporal variations of the NIRCam PSF are significant:
the left panel shows the residuals we added to the library ePSF
calculated iteratively modelling and subtracting the stellar models on
a grid of $5 \times 5$ residual points. On the right we illustrate how
the quality-of-fit (qfit) diagnostic\footnote{The quality-of-fit
  parameter describes the mean difference between the real star and
  the model, and it is defined as qfit$= \sum_{i,j}{(P_{i,j}-f_{\star}
    \times \psi )} / \sum_{i,j}{P_{i,j}}$, where $i=x_0-5, ...,
  x_0+5$, $j=y_0-5, ..., y_0+5$, $(x_0,y_0)$ is the star centre,
  $P_{i,j}$ is the value of the pixel in ($i,j$), $f_\star$ is the
  total flux of the star from the PSF-fitting, and $\psi$ is the PSF
  model. }, which assess the quality of a PSF model fit to an
individual star (the smaller is the qfit, the better the PSF fit),
improves significantly after the perturbation of the library ePSFs.

We conclude that, when possible based on the number of available
bright stars, perturbation of the library ePSFs such as those
presented here are necessary to achieve the best photometric (and
presumably astrometric) measurements.

\subsection{M~92 data reduction}

With the library of ePSF model arrays determined for the full set of
filters and detectors in our data-set, we extracted the positions and
fluxes of stars in each individual image adopting the PSF-fitting
software \texttt{img2xym}, already used and described in
Sect.~\ref{sect:mastercat}.  We set a faint limit for source detection
(local maxima above the neighbouring sky) as an integrated flux of at
least 50 counts, and at least five pixels from the closest source.

We cross-matched and transformed all source positions onto a common
reference system to derive a final catalogue of stars across each
image dither set.  Because the exposures are dithered by only a few
tens of pixels, individual detectors do not overlap. We therefore
analysed detector and filter combinations individually, generating
separate catalogues for each.  Sources were required to be detected in
at least three images.  Each catalogue includes the reference pixel
positions of the stars, their instrumental magnitudes\footnote{The
  instrumental magnitude is defined as $m_{\rm INSTR} = -2.5 \times
  \log{{\rm f_{\rm PSF}}}$, where ${\rm f_{\rm PSF}}$ is the total
  flux of the star (in counts) measured through PSF-fitting.} and
uncertainties, and the mean qfit.

We purged these catalogues of bad detection, PSF artefacts, and other
contaminants by using the procedures described in
\citet{2018MNRAS.481.3382N} and illustrated in Fig.~\ref{fig:4}.
Specifically, we used both photometric errors as a function of
instrumental magnitude and the qfit statistic to identify and remove
non-stellar sources. Figure~\ref{fig:4} also shows the instrumental
F090W versus F090W-F150W CMD before and after this cleaning procedure.
It is important to note that this CMD includes sources not only
clearly detectable in individual images, but present in at least 3 out
of 4 images. As such, these preliminary CMDs do not go as faint as the
data would allow, and are probably rather incomplete.  Moreover,
derived photometry are still vulnerable to pixel-area effects, these
due to geometric distortion, to lithographic effects (e.g.,
\citealt{1999PASP..111.1095A,2011PASP..123..622B}), or to both.
Future work will address these issues.

\begin{figure}
\centering
\includegraphics[bb=63 400 288 695, width=0.4\textwidth]{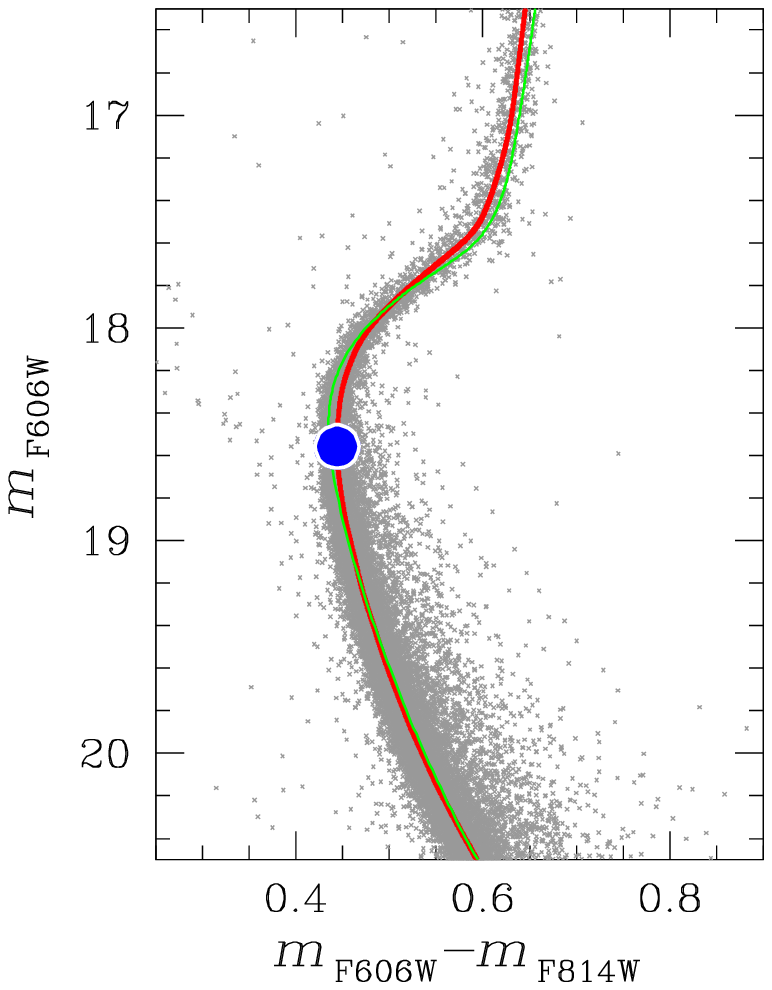}
\caption{The Main Sequence Turnoff region (blue point) of M~92 in the
  {\it HST} $m_{\rm F606W}$ versus $m_{\rm F606W}-m_{\rm F814W}$
  CMD. The fiducial line is plotted in red, while the green line
  traces the 13~Gyr BASTI-IAC isochrone, shifted to match the observed
  CMD by adjusting the distance modulus and reddening as determined in
  this work (see text for details). \label{fig:5}}
\end{figure}

\begin{figure}
\includegraphics[bb=34 265 325 691, width=0.4\textwidth]{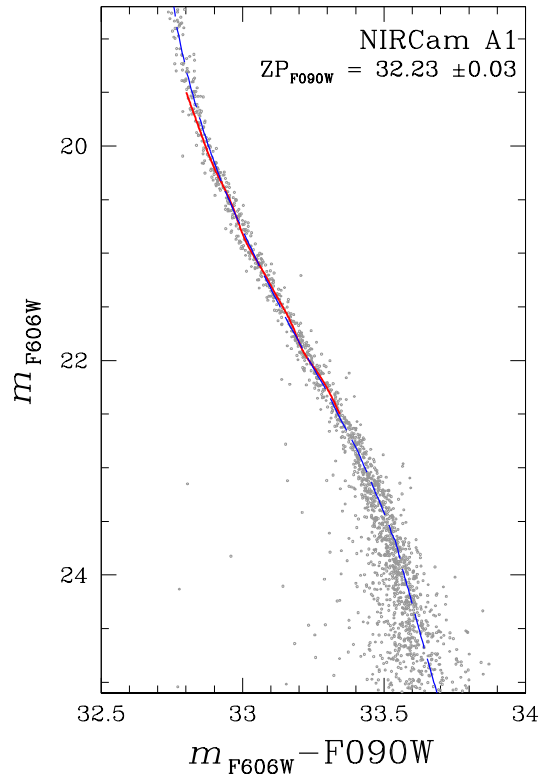}
\caption{Photometric calibration of data from filter F090W of detector
  1 of module A in the SW channel. Gray points are the $m_{\rm F606W}$
  versus $m_{\rm F606W}-$F090W CMD (where F090W is uncalibrated
  instrumental magnitudes). The red line represents the fiducial line
  of the CMD. The dashed blue line is the 13~Gyr BASTI-IAC isochrone,
  shifted by the distance modulus and reddening values determined in
  this work, and offset by the desired F090W photometric
  zero-point. \label{fig:6}}
\end{figure}

\begin{table*}
\caption{Photometric zero-points to bring instrumental magnitudes from
  different detectors onto a common photometric reference system
  consistent with models by BASTI-IAC for the assumed cluster
  parameters determined from optical \textit{HST} data.}
\resizebox{0.95\textwidth}{!}{ \begin{tabular}{l c c l c c }
\hline
\hline
\multicolumn{6}{c}{SW Channel} \\
\hline
{\bf Detector } & {\bf F090W} & {\bf F150W} &  {\bf Detector} & {\bf F090W} & {\bf F150W} \\
A.1 & $32.23 \pm  0.03$  & $31.88 \pm  0.03$ & B.1  & $32.10 \pm  0.03$ &  $31.84 \pm  0.04$ \\
A.2 & $32.20 \pm  0.04$  & $31.89 \pm  0.04$ & B.2  & $32.08 \pm  0.04$ &  $31.83 \pm  0.04$ \\
A.3 & $32.20 \pm  0.04$  & $31.86 \pm  0.05$ & B.3  & $32.11 \pm  0.05$ &  $31.85 \pm  0.05$ \\ 
A.4 & $32.28 \pm  0.05$  & $31.98 \pm  0.06$ & B.4  & $32.17 \pm  0.06$ &  $31.95 \pm  0.06$ \\ 
\hline
\hline
\multicolumn{6}{c}{LW Channel} \\
\hline
{\bf Module } & {\bf F277W} & {\bf F444W} &  {\bf Module} & {\bf F277W} & {\bf F444W} \\
A   & $31.01 \pm  0.07$  & $30.28 \pm  0.07$ & B  & $31.12 \pm  0.06$ &  $30.34 \pm  0.07$ \\
\hline
\end{tabular}
 }
\label{tab:1}
\end{table*}

\begin{figure*}
\includegraphics[bb=19 241 525 707, width=0.95\textwidth]{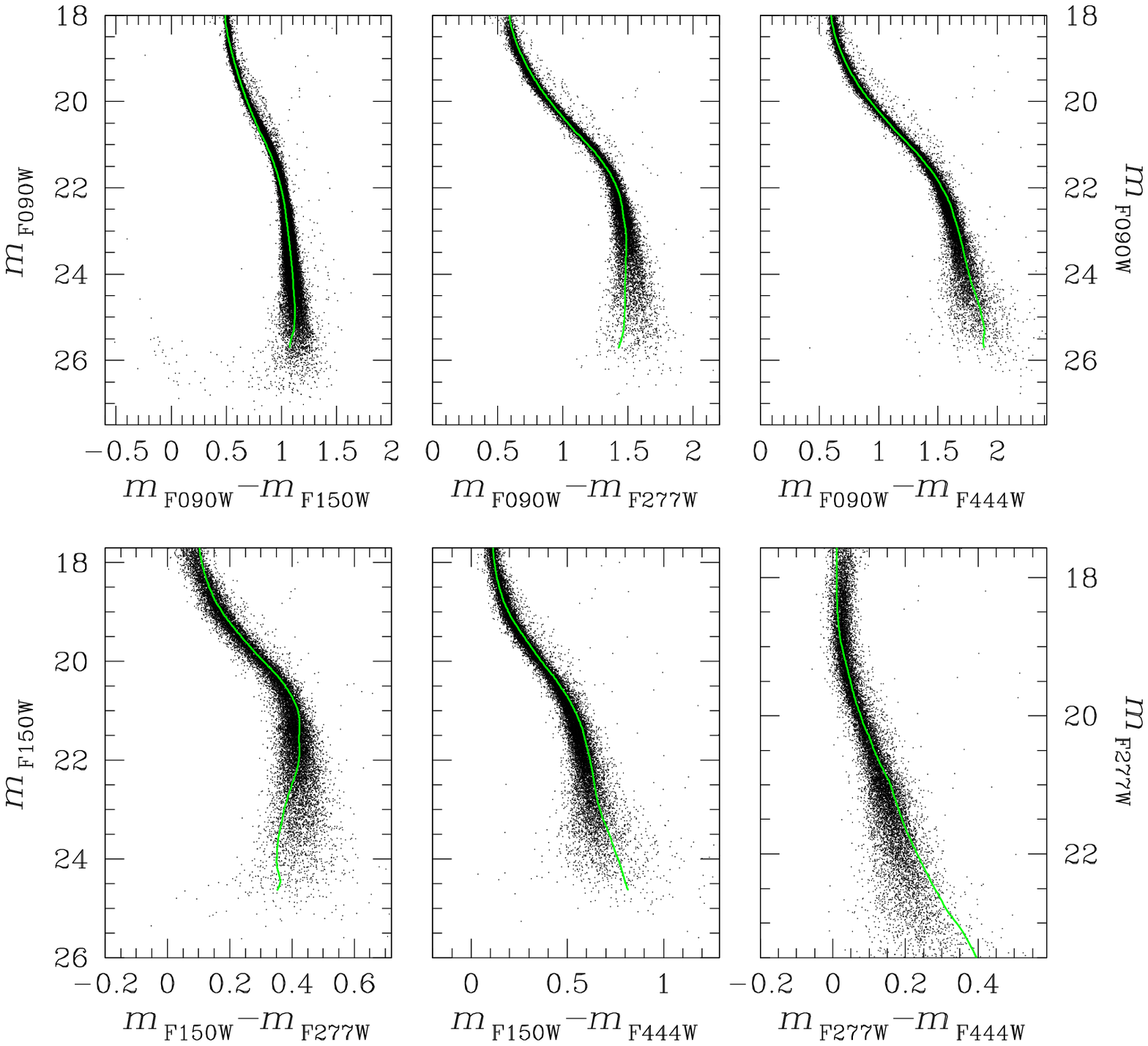}
\caption{Colour-magnitude diagrams inferred from the calibrated \textit{JWST} data. The green lines are the BASTI-IAC 13~Gyr
  isochrone. \label{fig:7}}
\end{figure*}

\section{Photometric registration}
\label{sec:cal}

As a first attempt at calibrating the instrumental magnitudes of the
stars detected in the M~92 observations, we cross-matched our
catalogues with the \texttt{\_cat} outputs of the Stage 3 pipeline
\texttt{calwebb\_image3}, based on pre-flight data.  However, when
comparing the CMDs obtained for stars on different detectors and
modules, it was clear that there were significant photometric
zero-point offsets, up to 0.4 mag.  There were also significant
deviations in overlapping CMDs obtained matching the \textit{JWST}
\texttt{\_cat} catalogues with the publicly available {\it HST}
catalogue of M~92 released by the "HST UV Globular cluster Survey"
(HUGS\footnote{\url{https://archive.stsci.edu/prepds/hugs/}},
\citealt{2018MNRAS.481.3382N}), based on data collected under {\it
  HST} programmes GO-10775 (PI: Sarajedini,
\citealt{2007AJ....133.1658S}) and GO-13297 (PI: Piotto,
\citealt{2015AJ....149...91P}).  
These photometric offsets have been also observed by other
  teams (see for example the research note by
  \citealt{2022arXiv220903348B}, or the zero-point corrections
  tabulated by
  G.~Brammer\footnote{\url{https://github.com/gbrammer/grizli/pull/107}}). The
  photometric offsets are correlated to the absolute flux zero-points adopted
  by the JWST pipeline to calibrate the single detectors. Actually,
  the JWST calibration programmes are still ongoing, and the
  zero-points used in the calibration pipeline are still far to be
  perfect.
We therefore pursued an alternative approach, boostrapping the {\it
  JWST} measurements to the {\it HST} data and theoretical models.

We first measured the distance of M~92 by comparing the {\it HST}
$m_{\rm F606W}$ versus $m_{\rm F606W}-m_{\rm F814W}$ CMD with the
BASTI-IAC models \citep{2018ApJ...856..125H,2021ApJ...908..102P}.  For
the models, we assumed an age $\tau=13.0$~Gyr, metallicity
[Fe/H]=$-2.31$, and alpha enrichment [$\alpha$/Fe]=+0.4
\citep{1996AJ....112.1487H}.  We selected the stars in magnitude
interval $16.5 \le m_{\rm F606W} \le 20.5$, roughly $\pm 2 $
magnitudes around the main sequence (MS) turn-off (TO). We determined
the fiducial line of these stars by using a naive estimator. We
divided the MS and red giant branch (RGB) sequences into bins of
$\delta m_{\rm F606W} = 0.5$ mag, and in these intervals defined a
grid of $N$ points separated by steps of width $\delta m_{\rm
  F606W}/4$. In each interval $m^i_{\rm F606W} < m_{\rm F606W} <
m^{i+1}_{\rm F606W}$, with $i=1,...,N$, we calculated the median
colour and magnitude of the stars within the magnitude interval. We
then interpolated these median points using a cubic spline (for a more
detailed description of this procedure, see
\citealt{2015MNRAS.451..312N, 2021A&A...646A.125L}). We then
calculated the $\chi^2$ between the fiducial line and the isochrone
model, shifted by a grid of distance modulii and reddening values, and
determined a minimum $\chi^2$ for a distance modulus $(m-M)_0=14.65
\pm 0.07$ ($d=8.51 \pm 0.28$~Kpc) and reddening $E(B-V)=0.02 \pm 0.01$
(see Fig.~\ref{fig:5}).  These values are in agreement with
measurements previously reported in the literature (
\citealt{2021MNRAS.505.5957B} and references therein).

To register the photometry of stars in the {\it JWST} catalogues, we
cross-matched these with sources in the {\it HST} HUGS catalogue of
M~92.  We then generated $m_{\rm F606W}$ versus $m_{\rm F606W}-X$
CMDs, where the $X$ represents uncalibrated photometry in each of the
NIRCam filters F090W, F150W, F227W, and F444W.  We then calculated
fiducial lines in a range of magnitudes $19.5 \le m_{\rm F606W} \le
22.5$, along the MS, following the procedure described above.
Shifting the $m_{\rm F606W}$ versus $m_{\rm F606W}-X$ model isochrone
onto the observational plane using the distance modulus and reddening
computed above, we calculated the mean difference between the colour
of the fiducial line and the shifted isochrone.
This value represents the photometric zero-point needed to bring
instrumental magnitudes from each detectors onto a common photometric
VEGAMAG reference system defined by the BASTI-IAC models for
the assumed cluster parameters (reddening, distance, [Fe/H],
[$\alpha$/Fe], etc).  The registration procedure is illustrated in
Fig.~\ref{fig:6} for the case of F090W data from detector 1 of module
A in the SW channel. Table~\ref{tab:1} summarises the zero-points we
determined for all of the detectors and filters, with uncertainties
taking into account contributions from the uncertainties on the
photometry, reddening, and distance modulus. The offsets
  between the different photometric zero-points are in agreement,
  within the errors, with those tabulated online by G.~Brammer.

For channel LW, which has only a single detector per module, we
evaluated the possibility of spatial variation of the photometric
zero-point across the field of view of each module as follows.  We
first divided each module into four regions of
1024~pixel$\times$1024~pixel, and matched the F277W and F444W
catalogues corresponding to these regions to the F090W
catalogues.\footnote{We could not use the {\it HST} catalogues for
  this correction because about half of the field of view of each
  module is outside the field of view of the {\it HST} images} We then
calculated zero-point corrections for each region by comparing the
theoretical models and fiducial lines as described above.  We found
that the zero-point correction does vary by $\sim\pm0.05$ mag between
regions, indicating that such a correction is necessary for high
precision photometry.  This is approximately the same order of
variation among the zero-points in each of the individual SW detectors
within a given module.

\begin{figure*}
\includegraphics[bb=21 289 551 700, width=0.9\textwidth]{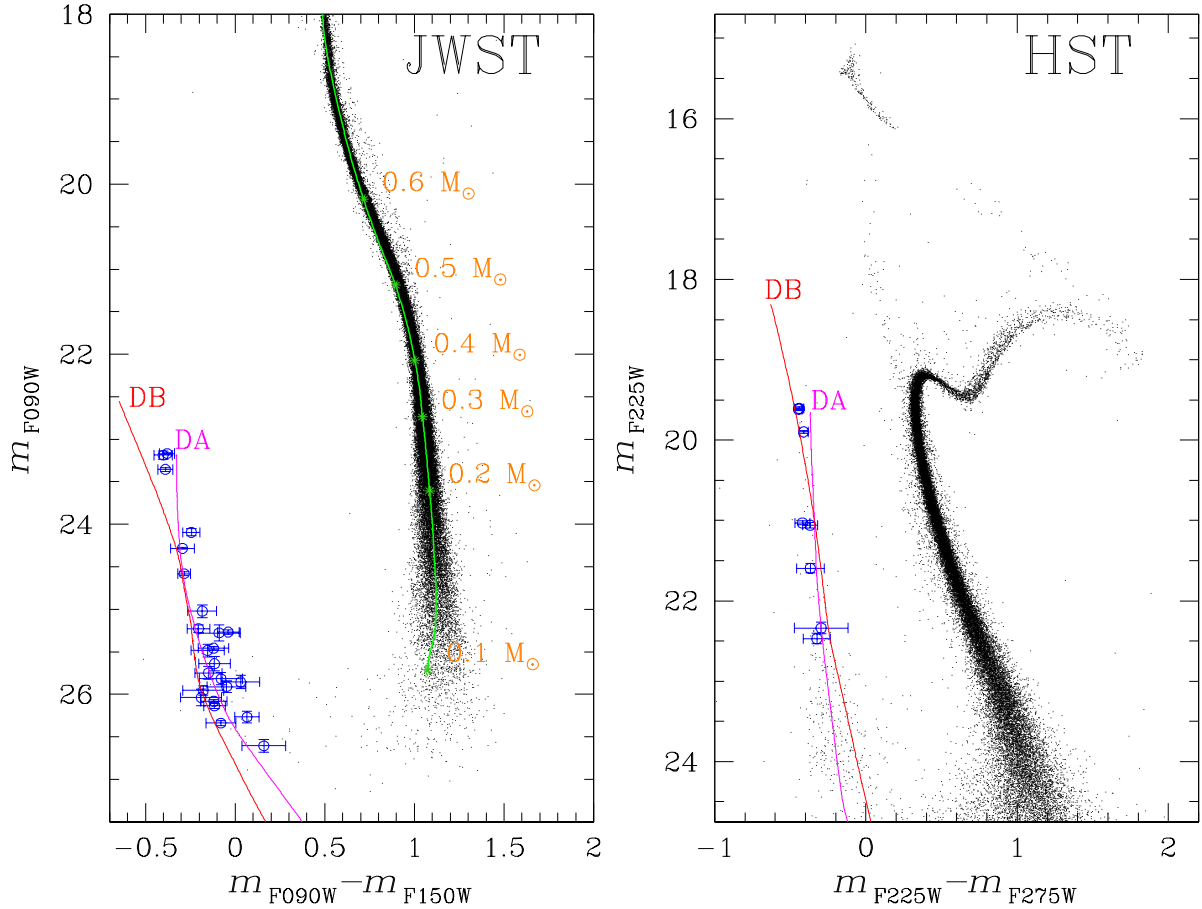}
\caption{Comparison between IR \textit{JWST} CMD ($m_{\rm F090W}$
  versus $m_{\rm F090W}-m_{\rm F150W}$; left) and UV {\it HST} CMD
  ($m_{\rm F225W}$ versus $m_{\rm F225W}-m_{\rm F275W}$; right).  The
  green line in the left panel is the BASTI-IAC isochrone used to
  derive the photometric zero-point, with masses in steps of
  0.1~$M_\odot$ are indicated. In both CMDs, blue circles indicate the
  WDs identified in \textit{JWST} data.  Magenta and red lines
  delineate WD cooling tracks for hydrogen and helium envelopes
  respectively. \label{fig:8}}
\end{figure*}

\section{Colour-magnitude diagrams of M\,92}
\label{sec:cmd}

With the individual zero-points calibrated, we can now evaluate the
infrared CMDs of M~92 revealed by the {\it JWST}
data.  Figure~\ref{fig:7} compares six CMD combinations derived by
combining all the \textit{JWST} catalogues extracted in this work.
We superimposed to these CMDs the BASTI-IAC isochrone models used to
calibrate the zero-points.  We find that all of the CMDs reach down to
or below the faint end of the MS locus, at least to a stellar mass of
about 0.1~$M_{\odot}$.  The CMD that reaches the bottom of the MS with
the highest signal-to-noise ratio (S/N) is $m_{\rm F090W}$ versus
$m_{\rm F090W}-m_{\rm F150W}$. Indeed, at $m_{\rm F090W} \sim 26$, we
measure a S/N $\sim 12$ in F090W and S/N $\sim 20$ in F150W, while for
stars at the equivalent F090W magnitude the LW measurements have S/N
$\sim 10$ for F277W and S/N $\sim 8$ for F444W.

\subsection{M92 White dwarfs}

In the $m_{\rm F090W}$ versus $m_{\rm F090W}-m_{\rm F150W}$ CMD we
identified 24 sources that passed the quality selection criteria at
magnitudes $m_{\rm F090W} \gtrsim 23$ and colour $(m_{\rm
  F090W}-m_{\rm F150W})\lesssim 0.3$.  We visually confirmed that
these sources correspond to robust point-sources in the stacked F090W
and F150W frames (Fig.~\ref{fig:ap1} in Appendix~\ref{sec:ap1}).
These stars are likely white dwarfs (WDs) in M~92. To verify this
hypothesis, we first cross-matched the F090W and F150W catalogues with
F225W and F275W UV catalogues generated by reducing {\it HST} data
collected in programme GO-15173 (PI: Kalirai). Figure~\ref{fig:8} displays
the resulting CMDs.  We found seven WD candidates were matched between
\textit{JWST} and \textit{HST} data sets, and all of them lie on the
WD sequence in the $m_{\rm F225W}$ versus $m_{\rm F225W}-m_{\rm
  F275W}$ CMD.  We also compared both IR and UV CMDs of the WD cooling
sequence with two 0.54$M_{\odot}$ cooling tracks with hydrogen and
helium atmospheres from the BaSTI-IAC database
\citep{2022MNRAS.509.5197S}.  We are observing the bright part of the
WD cooling sequence of M~92 in the {\it JWST} catalogues, where the mass
of the evolving WDs is practically constant and equal to $\sim 0.53
\pm 0.01 M_{\odot}$ \citep{2009ApJ...705..408K}.  The excellent
alignment of the selected sources with the theoretical cooling tracks
in both IR and UV CMDs confirms their WD nature.
Paper\,II will calibrate with the geometric distortion of NIRCam,
making possible proper motions measurements for those sources in
common with \textit{HST} collected between 3 and 20 years earlier,
which will confirm or refute the membership of these WDs in M~92.

\begin{figure*}
\includegraphics[bb=18 147 588 710, width=0.95\textwidth]{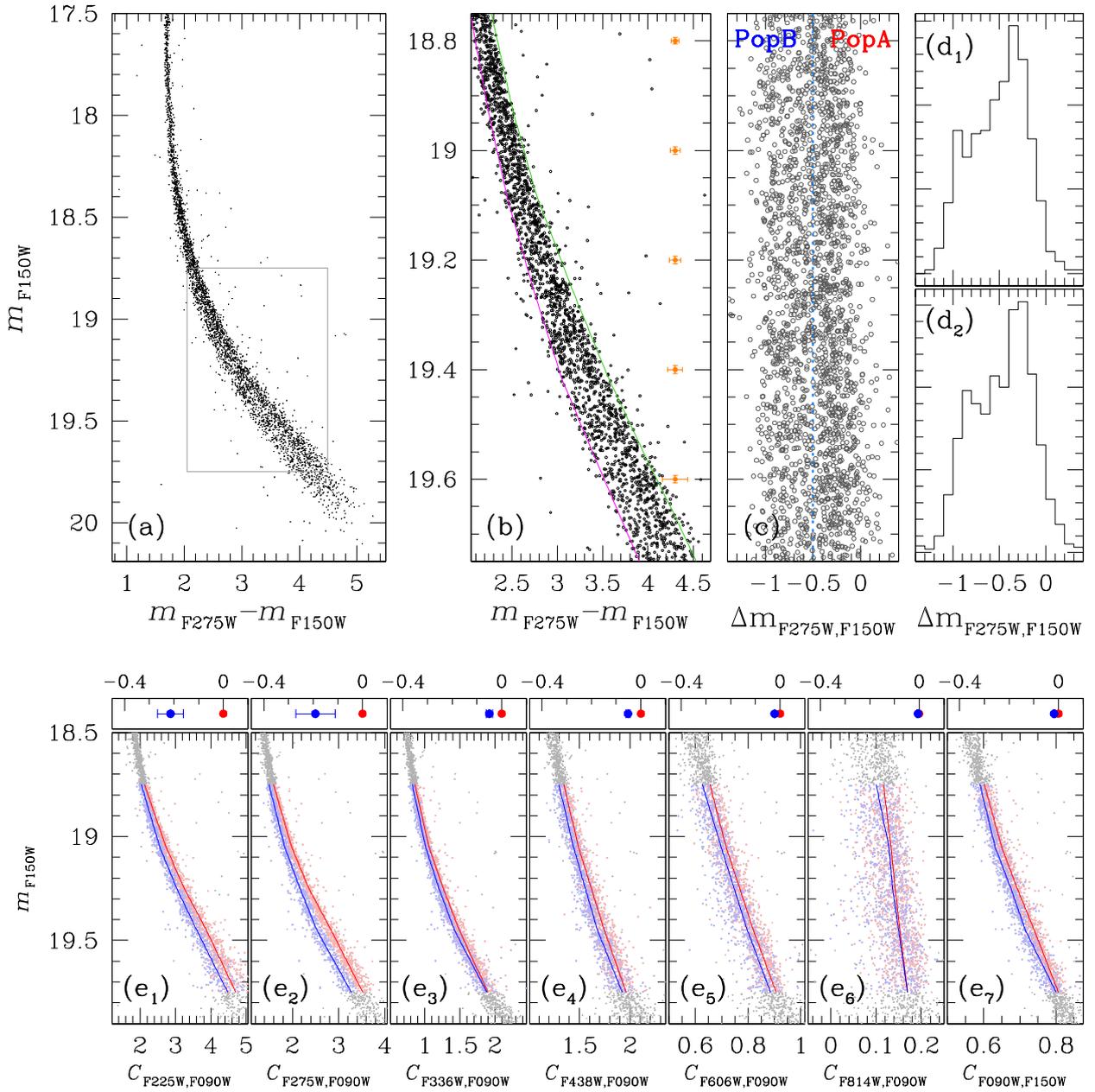}
\caption{Multiple stellar populations discerned in the UV/optical/IR
  CMDs of M~92. Panel (a) shows the $m_{\rm F150W}$ versus $m_{\rm
    F275W}-m_{\rm F150W}$ CMD where the spread of the MS is evident
  for $m_{\rm F150W} \gtrsim 18.75$. Panel (b) is a zoom of the MS for
  magnitudes $ 18.75 \leq m_{\rm F150W}\leq 19.75$; magenta and green
  lines display the 5th- and the 95th-percentile fiducial
  lines. Orange error bars along the right side of this panel indicate
  the photometric errors in intervals of 0.2 mag in F150W. Panel (c)
  shows the verticalized MS from panel (b).  We are able to visually
  divided the stars in two groups: PopA (right) and PopB
  (left). Panels (d$_1$) and (d$_2$) show the distribution of sources
  across the verticalized MS in the magnitude ranges $ 18.75 \leq
  m_{\rm F150W} < 19.25$ (panel (d$_1$)) and $ 19.25 \leq m_{\rm
    F150W} < 19.75$ (panel (d$_2$)). Panels (e$_1$) through (e$_7$)
  show the colour distributions of PopA and PopB stars selected from
  the verticalized MS in the $m_{\rm F150W}$ versus $C_{\rm
    X,F090W}=(m_{\rm X}-m_{\rm F090W})$ CMDs. Above each panel, the
  mean difference in colour between PopA (red) and PopB (blue) is
  indicated. \label{fig:10}}
\end{figure*}

\subsection{Multiple populations in M~92}

Like almost all the old Galactic globular clusters (e.g.,
\citealt{2015AJ....149...91P,2018ARA&A..56...83B}) and extragalactic
globular clusters (e.g.,
\citealt{2014ApJ...797...15L,2019MNRAS.485.3076N}), M~92 hosts
multiple stellar populations characterised by different light element
(C, N, O, Al, Mg, etc) and helium abundances.  Analysis by
\citet{2017MNRAS.464.3636M} indicates that M~92 hosts at least three
different stellar populations: a ``first generation'' (1P) having
primordial helium content and chemical properties similar to the
medium out of which it formed; and ``second generation'' populations
(2Ps) which are enriched in helium and nitrogen, and depleted in
carbon and oxygen, likely formed from the material processed by the
1P. \citet{2018MNRAS.481.5098M} found a maximum difference in helium
between the 1P and 2Ps populations of $\delta Y_{\rm max} \sim 0.04$.
\citet{2015AJ....149..153M} reported average values of [O/Fe]=0.67,
      [C/Fe]=$-$0.42, and [N/Fe]=0.89 for 1P, and [O/Fe]=0.40,
      [C/Fe]=$-$0.38, and [N/Fe]=1.03 for 2Ps.

While a detailed analysis of the multiple population phenomenon is
behind the scope of this work, we checked that the CMDs obtained with
the {\it HST} UV/optical and \textit{JWST} IR cross-matched
catalogues.
As shown by \citet{2019A&A...629A..40S}, the magnitudes of RGB and
cool MS stars in the \textit{JWST} filters contain signatures of CNO
abundance variations typical of the multiple population phenomenon.
Figure~\ref{fig:10} displays the $m_{\rm F150W}$ versus $m_{\rm
  F275W}-m_{\rm F150W}$ CMD, which shows a clear MS spread for $m_{\rm
  F150W}\gtrsim 18.75$. We analysed in detail this spread to determine
if it is due to photometric errors or the presence of multiple
populations. We first verticalized the MS in the magnitude range
$18.75\leq m_{\rm F150W}\leq 19.75$ calculating the 5th and 95th
percentile of the colour distribution along the MS.  We used these two
fiducial lines (fid(5th) and fid(95th)) to calculate the quantity:
\begin{equation}
\Delta m_{\rm F275W,F150W} = \frac{(m_{\rm  F275W}-m_{\rm F150W})-{\rm fid(95th)}}{{\rm fid(95th)}-{\rm fid(5th)}} .
\end{equation}
This verticalized MS is shown in Fig.~\ref{fig:10}, where the presence
of multiple sequences is visually evident.  We also discern multiple
peaks in distributions of stars across the verticalized MS. We
therefore divided the CMD population into two parts: we assigned
points with $\Delta m_{\rm F275W,F150W}>-0.5$ to a group called PopA,
and points with $\Delta m_{\rm F275W,F150W}\le -0.5$ to a group called
PopB.  PopA should contain the large part of 1P stars, so on average
its photometric properties reflect 1P's characteristics; while PopB
tracks the (unresolved) 2Ps populations We also evaluated the colour
distributions of these two groups of stars in a sequence of CMDs of
$m_{\rm F150W}$ versus $m_{\rm X}-m_{\rm F090W}$, with X = F225W,
F275W, F336W, F438W, F606W, F814W, and F150W ({\it HST} and {\it JWST}
filters), i.e. using colours different from that used for the
identification of the PopA/PopB groups in Fig.~\ref{fig:10}.  We found
that the two groups of stars form mostly separated sequences in each
of these CMDs, differing in colour by up to 0.2~mag. The consistent
separation of these groups strengthens the case that PopA and PopB
stars belong to populations with different chemical properties.

\section{Summary}
\label{sec:sum}
With the \textit{James Webb Space Telescope} now fully operational, a
new era of astronomy has begun. Its various instruments will allow us
to observe not only the primordial Universe and exoplanet atmospheres,
but also low mass stars in globular clusters and resolved galaxies.

In this work, we have evaluated \textit{JWST} data of the globular
cluster M~92 collected with the NIRCam instrument for DD-ERS programme
ERS-1334. We have used these data to derive a library of
spatially-varying effective PSFs for filters F090W/F150W (SW) and
F277W/F444W (LW) for each of the individual detectors and modules.
We have demonstrated that significant temporal variations are present
in NIRCam on the order of 3-4~\% based on a second set of observations
of the Wolf-Lundmark-Melotte irregular galaxy obtained one month later
under the same programme.  We have also shown that these variations can
be corrected by perturbing these library ePSFs, enabling
high-precision photometry for various crowded regions over time.
We have also demonstrated a method for calibrating instrumental
magnitudes for individual NIRCam detectors and filters using
theoretical models and prior {\it HST} observations of M~92, and
provide measurements of the photometric zero-points for the individual
detectors and filters used in this data-set.

Our analysis allows for the first examination of \textit{JWST}
infrared CMDs for a globular cluster.  We demonstrate that all of the
CMDs are well-fit to theoretical 13~Gyr isochrones, and reach the
lower bottom of M~92's Main Sequence at a S/N$\sim$ 10-20 depending on
the filter set.  We also detect white dwarfs in the CMD in the F090W
and F150W filters, and a cross-match of the \textit{JWST} catalogues
with the {\it HST} UV catalogue in F225W and F275W filters (obtained
by us reducing data from GO-15173, PI: Kalirai) identifies 7 WDs in
common. The location of the WDs on the CMDs are also in agreement with
theoretical $0.54~M_{\odot}$ WD cooling tracks.

We also searched for the presence of 
multiple stellar populations in the UV/IR CMD of the Main Sequence of
M~92.  We demonstrated that the $m_{\rm F150W}$ versus $(m_{\rm
  F275W}-m_{\rm F150W})$ CMD resolves a spread of the MS in the mass
range of 0.5-0.6~$M_{\odot}$ that is not attributable to photometric
errors.  Dividing the MS into two groups, PopA and PopB, corresponding
approximately to 1P and 2P stars, respectively, we analysed the
distributions of these populations in different combinations of
colours.  We found that the two groups almost always form two
well-separated sequences, confirming that they are composed of stars
with different physical and chemical properties.

\section*{Acknowledgements}
The authors warmly thank the anonymous referee for carefully
reading the paper and for the useful comments and suggestions that have
contributed to improving the quality of the manuscript.
The authors DN, LRB, MG and MS acknowledge support by MIUR under PRIN program \#2017Z2HSMF and 
by PRIN-INAF\,2019 under program \#10-Bedin.
MS acknowledges support from The Science and Technology Facilities
Council Consolidated Grant ST/V00087X/1. SC acknowledges financial
support from \lq{Progetto Premiale}\rq\ MIUR {\sl MITIC} (PI:
B. Garilli), progetto INAF Mainstream (PI: S. Cassisi) and from {\em
    PLATO} ASI-INAF agreement n.2015-019-R.1-2018, and from INFN
  (Iniziativa specifica TAsP).
This work is based on observations made with the NASA/ESA/CSA James
Webb Space Telescope. The data were obtained from the Mikulski Archive
for Space Telescopes at the Space Telescope Science Institute, which
is operated by the Association of Universities for Research in
Astronomy, Inc., under NASA contract NAS 5-03127 for \textit{JWST}. These
observations are associated with program ERS-1334 (PI: Weisz).  
The authors acknowledge the ``The \textit{JWST} Early Release Science Program
for Resolved Stellar Populations'' team led by D.~Weisz, J.~Anderson,
M.~L.~Boyer, A.~A.~Cole, A.~E.~Dolphin, M.~C.~Geha, J.~Kalirai,
N.~Kallivayalil, K.~B.~W.~McQuinn, K.~M.~Sandstrom, B.~F.~Williams for
developing their observing program with a zero-exclusive-access
period.
This research is also based on observations made with the NASA/ESA
Hubble Space Telescope obtained from the Space Telescope Science
Institute, which is operated by the Association of Universities for
Research in Astronomy, Inc., under NASA contract NAS 5–26555. These
observations are associated with programs GO-10775 (PI: Sarajedini),
GO-13297 (PI: Piotto), GO-15173 (PI: Kalirai).

\section*{Data Availability}
The data underlying this article are publicly available in the Mikulski Archive
for Space Telescopes at \url{https://mast.stsci.edu/}. 
The catalogues underlying this work are available in the online supplementary material of the article.


\bibliographystyle{mnras}
\bibliography{biblio}

\appendix

\section{White dwarfs on the stacked image}

\begin{figure*}
\centering
\includegraphics[width=0.31\textwidth]{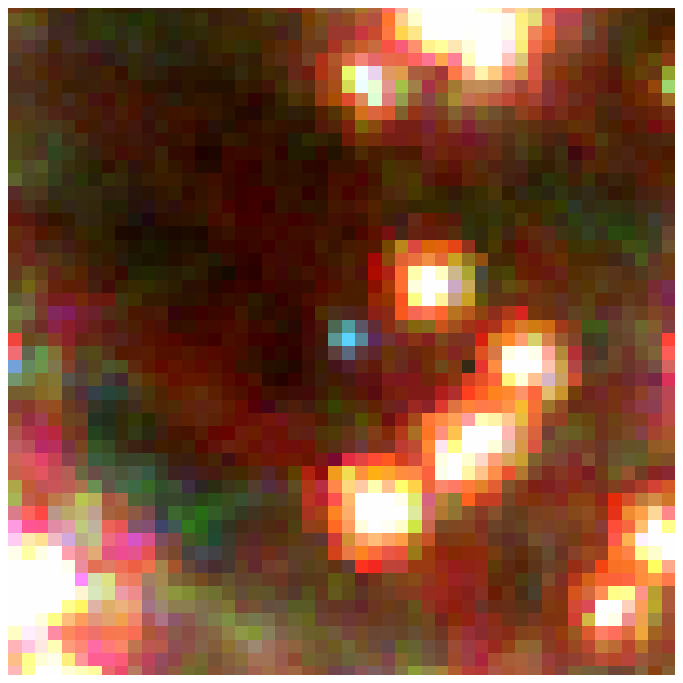}
\includegraphics[width=0.31\textwidth]{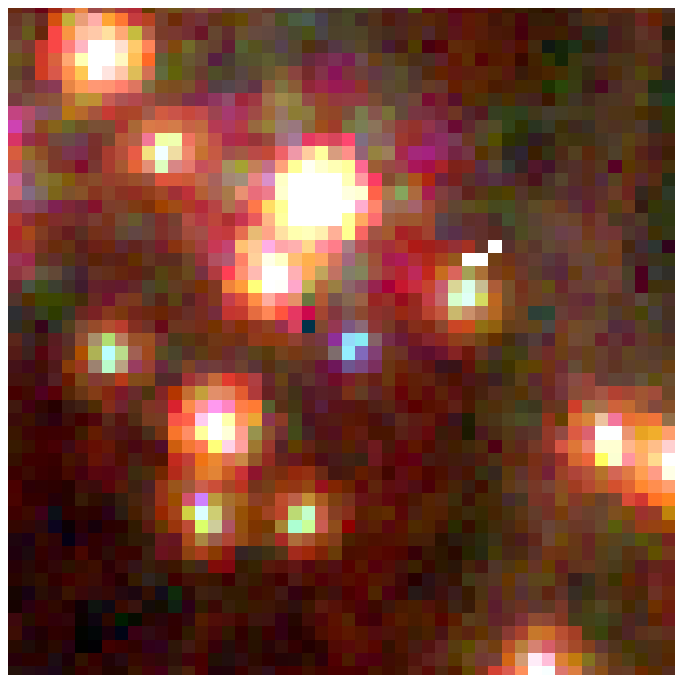}
\includegraphics[width=0.31\textwidth]{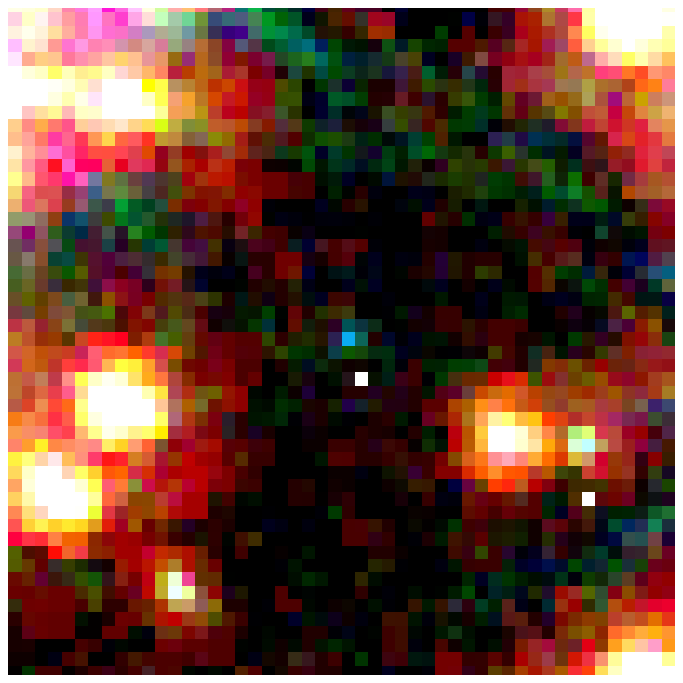}
\caption{Three white dwarfs on the  colour-stacked image.\label{fig:ap1}}
\end{figure*}

\label{sec:ap1}
In Figure~\ref{fig:ap1} we show three WDs on the three-colour stacked
image (F090W$+$F150W$+$F444W), to demonstrate that they are point
sources. The first two WDs from the left have magnitude $m_{\rm F090W}
\sim 23.2$, the last have magnitude $m_{\rm F090W} \sim 25.8$.

\section{Electronic material}
The catalogues obtained in this work will be available electronically
as supporting material to this paper. We will release a catalogue for
each detector and for each filter (totally 20 catalogues). A
description of the columns of the catalogues is reported in
Table~\ref{tab:a1}.  We will also release the PSF arrays on the
website \url{https://web.oapd.inaf.it/bedin/files/PAPERs_eMATERIALs/JWST/Paper_01/}

\begin{table*}
\caption{Description of the columns of the released catalogues.}
\begin{tabular}{l c c l}
\hline
Column  & Name  & Unit  & Explanation \\
\hline
01 & \texttt{X}  & pixel & X stellar position in the reference system of the detector \\
02 & \texttt{Y}  & pixel & Y stellar position in the reference system of the detector \\
03 & \texttt{MAG}& &Calibrated VEGAMAG magnitude  \\
04 & \texttt{SIGMA\_MAG}& & Photometric error \\
05 & \texttt{QFIT}& & Quality-fit parameter \\
\hline 
\end{tabular}
\vspace{1ex}

{\raggedright {\bf Note: } The name of each file reports the information on the catalogue: \texttt{NGC6341.JWST.X.YZ.CAT}, 
where \texttt{X}=F090W, F150W, F277W, F444W is the filter, \texttt{Y}=A, B is the module, and (only for SW channel) \texttt{Z}=1,2,3,4 is the detector number. \par}

\label{tab:a1}
\end{table*}

\label{lastpage}
\end{document}